\RequirePackage{ifpdf}
\documentclass{JHEP3}
\usepackage{graphics}
\usepackage[pdftex]{graphicx}
\usepackage{amsmath}
\usepackage{cite}
\usepackage{feynmp}
\def\be{\begin{equation}}
\def\ee{\end{equation}}
\def\bea{\begin{eqnarray}}
\def\eea{\end{eqnarray}}
\def\eps{\epsilon}

\def\eps{\epsilon}

\def\cO{  {\cal O}  }

\preprint{}
\title{Higher loop mixed correlators in N=4 SYM}

\author{Luis F.\ Alday$^{a}$, Johannes M.\ Henn$^{b}$, Jakub Sikorowski$^{c}$\\
$^{a}$ Mathematical Institute, University of Oxford, 
   Oxford OX1 3LB, U.K. \\
$^{b}$ Institute for Advanced Study, Princeton, NJ 08540, USA\\
$^{c}$ Rudolf Peierls Centre for Theoretical Physics, University of Oxford, Oxford OX1 3NP, U.K.\\

\email{alday@maths.ox.ac.uk ,\\ jmhenn@ias.edu, \\ j.sikorowski12@physics.ox.ac.uk}}

\abstract{
We compute analytically the two-loop contribution to the correlation function of the Lagrangian 
with a four-sided light-like (or null) Wilson loop in
  ${\cal N}=4$ super Yang-Mills.
As a non-trivial test of our result, we reproduce the three-loop value of the cusp anomalous dimension 
upon integration over the insertion point of the Lagrangian. 
The method we used involved calculating a dual scattering amplitude.
Moreover, we give a simple representation of the loop integrand of the latter in twistor variables.
}


\begin{document}

\section{Introduction and main results}

Over the last few years there has been remarkable progress in the computation of observables in planar ${\cal N}=4$ super Yang-Mills (SYM). These observables include the S-matrix, correlation functions of local operators, Wilson loops and combinations of these. An interesting class of observables is the correlation function of Wilson loops with local operators. In particular cases such correlators are fixed by the symmetries of the theory \cite{Berenstein:1998ij}, but in general they contain useful dynamical information. 

In this paper we consider the simplest correlators not fixed by symmetries: the correlation function of a polygonal light-like (or null) Wilson loop with four edges and a local operator, which we take to be the Lagrangian $\mathcal{L}$ of the theory \cite{Alday:2011ga,Engelund:2011fg}.
Such a correlation function has ultraviolet (UV) divergences characteristic of light-like Wilson loops \cite{Korchemskaya:1992je,Drummond:2007au}.
In order to obtain a finite observable, we normalize this correlation function by the same correlator without
the Lagrangian insertion.  The finiteness of the ratio follows from the structure of UV divergences of light-like Wilson loops. 
Indeed our perturbative results agree with this expectation.  

Conformal symmetry implies that the overall scaling dimension of this observable is fixed. Moreover, it is a non-trivial function of one kinematic cross-ratio only,
\begin{align}\label{definitionF}
\frac{ \langle W_{4}(x_1, x_2, x_3,x_4) {\mathcal L}(x_5) \rangle}{\langle W_{4}(x_1,x_2,x_3,x_4) \rangle}  = \frac{1}{\pi^2} \frac{x_{13}^2 x_{24}^2}{x_{15}^2 x_{25}^2 x_{35}^2 x_{45}^2} F(x) \,,
\end{align}
where $x_{ij}^2=(x_{i}-x_{j})^2$, and
\begin{align}
x= \frac{x_{25}^2 x_{45}^2 x_{13}^2}{x_{15}^2 x_{35}^2 x_{24}^2}
\end{align} 
is the only cross-ratio which can be formed by the locations of the cusps $x_{1,...,4}$ (subject to the 
conditions $x_{12}^2=x_{23}^2=x_{34}^2=x_{41}^2=0$), and 
the insertion point of the local operator $x_5$.
Note that as a consequence of the symmetry under cyclic permutations of $x_{1,...,4}$ , $F$ satisfies the symmetry property $F(x)=F(1/x)$. 

$F$ also depends on the rank of the gauge group $N$ and the `t Hooft coupling $\lambda = g^2 N$.
We will consider $F$ in the planar limit, where it has the perturbative expansion
\begin{align}
F(x) = \sum_{L=1}^{\infty} \left( \frac{\lambda}{8 \pi^2} \right)^L F^{(L-1)}(x) \,.
\end{align}
The first non-planar corrections can appear at four loops.
The tree-level and one-loop contributions to $F$ have been computed in  \cite{Alday:2011ga,Alday:2012hy}, with the result
\begin{align}
F^{(0)}(x) =& - \frac{1}{2} \,,\\
F^{(1)}(x) =& \frac{1}{4} \left[ \log^2 x + \pi^2 \right]  \label{Foneloop}\,. 
\end{align}
In this paper, we compute analytically the two-loop contribution. We obtain
\begin{align}\label{Ftwoloop}
F^{(2)}(x) =& -\frac{1}{8} \Bigg\{ \frac{1}{2} \log^4 x
 + \log^2 x \Big[ -\frac{2}{3} L_{2}(x) + 12 \zeta_2 \Big] 
+ \log x \Big[ -4 L_{3}(x)  \Big]  \nonumber \\
& 
\;\;\;\;\;\;+ \Big[ -\frac{2}{3} (L_2(x))^2 - 8 L_{4}(x) - 16 \zeta_2 L_{2}(x) +107 \zeta_4 \Big]
 \Bigg\} \,,
\end{align}
where the functions
\begin{align}
L_{n}(x) :=& {\rm Li}_{n}\left(\frac{1}{1+x}\right) +  \, {\rm Li}_{n}\left( \frac{x}{1+x} \right) -\zeta_n  \,,  \quad n \;\; {\rm even} \,, \\
L_{n}(x) :=& {\rm Li}_{n}\left(\frac{1}{1+x}\right) - \, {\rm Li}_{n}\left( \frac{x}{1+x} \right)  \,,  \quad n \;\; {\rm odd} \,.
\end{align}
are manifestly symmetric (antisymmetric) under $x \to 1/x$ for $n$ even (odd).

The correlation functions on the l.h.s. of eq. (\ref{definitionF}) can in principle be evaluated in configuration
space. However, there is also a dual formulation of the same objects in terms of integrals
resembling scattering amplitudes \cite{Alday:2010zy,Eden:2010zz}.
In order to see this, one can think about loop corrections to the 
correlation functions being generated by Lagrangian insertions. Formally, i.e. neglecting regulator issues,
one then has the integrand of an $(L+1)$-loop four-point scattering amplitude, with $L$ integrations to be carried out.
Since all divergences cancel in the ratio of eq. (\ref{definitionF}), one can argue that $F$ can be also obtained from a 
dual calculation, where both numerator and denominator in eq. (\ref{definitionF}) are replaced by four-point on-shell scattering amplitudes,
the numerator having an additional Lagrangian insertion. 
A representation of $F$ in terms of scattering amplitude integrals was given in ref. \cite{Alday:2012hy}.
Note that in this dual representation, UV divergences of the Wilson loop are transformed into infrared (IR) divergences
of scattering amplitudes. Of course, the latter cancel in the final result for $F$.

The integrals one obtains are those for the scattering of four massless particles, but with the unusual feature of involving
an operator insertion at the point $x_{5}$. If desired, one can use conformal symmetry of the Wilson loops (i.e. dual conformal
symmetry in the scattering amplitude picture) to send this point to infinity. 

We have performed the calculation both using dimensional regularization, as well as in a mass regularization setup \cite{Alday:2009zm}. 
The quantify $F$ is expected to be scheme independent, see \cite{Henn:2011by}, and indeed we verified (numerically) that
both calculations gave the same finite result.
We found that the calculation was simpler in the massive regularization, as expected based on previous
experience with similar integrals \cite{Drummond:2010mb,Dixon:2011nj}.

Over the last couple of years, twistor techniques have been tremendously successful in describing scattering amplitudes 
of ${\cal N}=4$ SYM at weak coupling, both at the level of the integrand, see e.g. \cite{Hodges:2009hk,Mason:2009qx,ArkaniHamed:2010kv,ArkaniHamed:2010gh},
and for obtaining analytic integrated expressions, \cite{Drummond:2010mb,Dixon:2011nj,Drummond:2010cz,Alday:2010jz}.
A natural question is whether such techniques will also be useful to understand correlation functions of local operators. 
As a step towards this, we give a simpler, twistorial representation of the integrand at two loops. 
This is closely related to similar simplifications observed when studying the exponentiation of scattering
amplitudes \cite{Drummond:2010mb}.
This representation has several advantages, as we shall discuss.
At the one-loop level, we see that the result can be written in terms of a single finite integral.
Being a one-loop integral, the latter is of course known. However, perhaps the simplest
way of obtaining this result is to derive a differential equation \cite{Drummond:2010cz} for this
single-variable function, which can be readily solved.
At two loops, we find a very compact representation in terms of five integrals, 
each of which has no subdivergences.
We show that the remaining overall divergence cancels between the different terms.
We find it likely that the differential equation technique of ref.  \cite{Drummond:2010cz} and related twistor-space methods will allow for a simpler evaluation of these integrals in the future.

Finally, another interesting feature of the above correlator is that by integrating over the point where the Lagrangian is inserted, we recover the expectation value of the four sided Wilson loop, and in particular, from its most divergent part, the light-like cusp anomalous dimension $\Gamma_{\rm cusp}$ \cite{Korchemskaya:1992je}. 
Denoting the necessary infrared regulator by $\Lambda$, one obtains a formula (schematically)
\begin{align}
\int_{\Lambda} \frac{d^{4}x_{5}}{i \pi^2} \frac{F(x)}{\prod_{j=1}^{4} x_{j5}^2}  \sim  \lambda \frac{\partial}{\partial \lambda} \Gamma_{\rm cusp}  \, \log^2  \Lambda  + \cO(\log \Lambda)\,.
\end{align} 
Regulator subtleties and the precise form of this identity for two different regularizations will be discussed in the body of the paper. 
This allows to extract the cusp anomalous dimension from a finite quantity. 
Conversely, knowing the cusp anomalous dimension independently, we obtain an integral constraint on the result. We check that the perturbative results up to two loops as well as the strong coupling result, give rise to the correct value of the cusp anomalous dimension.

The outline of this paper is as follows. In section two we describe how to obtain the analytic result at two loops from the integral representation previously found in \cite{Alday:2012hy}. 
In section three we give a twistorial representation for the loop integrals and in section four we show that by integrating over the insertion point of the local operator we reproduce the correct value of the cusp anomalous dimension. We end up with a summary of our results and outlook, while several technical points are relegated to the appendices.

\section{Analytic two-loop calculation from loop integrals}\label{section:two-loop_calculation}

\subsection{Expression in terms of loop integrals}
The expression for $F$ in terms of loop integrals has been written out in ref. \cite{Alday:2012hy}.
Converting to more standard conventions for Minkowski-space loop integrals, we have
\begin{align}\label{Fintegrals}
F(x) =& -\frac{1}{2} \left( \frac{\lambda}{8 \pi^2} \right)  \nonumber \\
   & + \frac{1}{4}  \left( \frac{\lambda}{8 \pi^2} \right)^2 \left[ F_{1235} + F_{4125} + F_{3415} + F_{2345} -F_{1234} \right] \nonumber
    \\
    & - \frac{1}{8}  \left( \frac{\lambda}{8 \pi^2} \right)^3 \sum_{8\; {\rm perm}} \left[ - \frac{1}{4} I_{1} + \frac{1}{2} I_2 + \frac{1}{2}  I_3 +  I_4 +\frac{1}{8} I_5 - \frac{1}{2} I_6 + \frac{1}{4}  I_7 \right] \,.
\end{align}
Here 
the $8$ permutations refer to $4$ cyclic permutations of the points $x_{1},x_{2},x_{3},x_{4}$, and to the swap $x_{1} \leftrightarrow x_{4}, x_{2} \leftrightarrow x_{3}$.

\begin{figure}[h!]
  \centering
  \def\svgwidth{2cm} 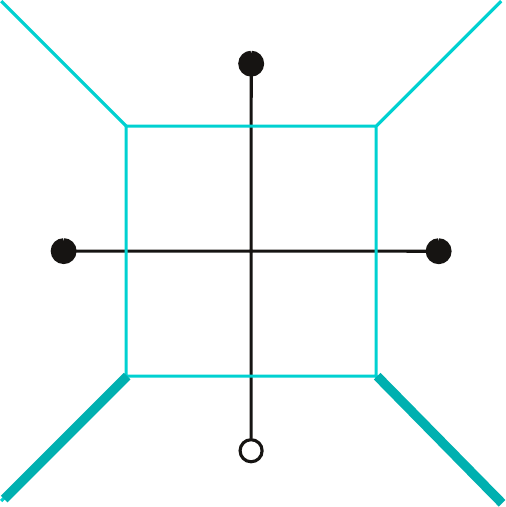 \qquad 
  \def\svgwidth{2cm} 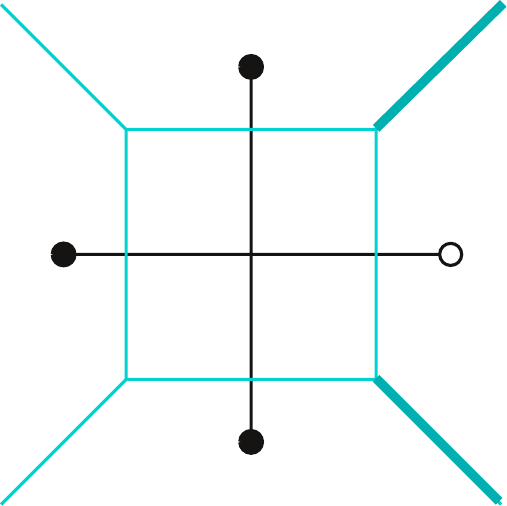 \qquad 
  \def\svgwidth{2cm} 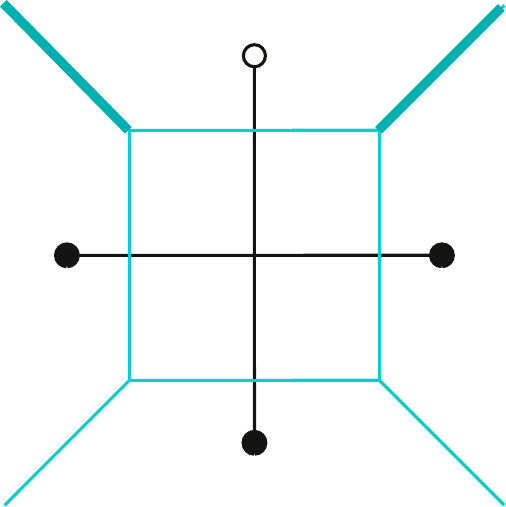 \qquad 
  \def\svgwidth{2cm} 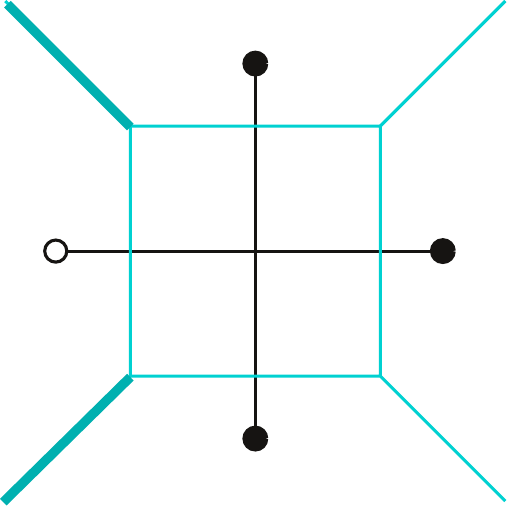 \qquad 
  \def\svgwidth{2cm} 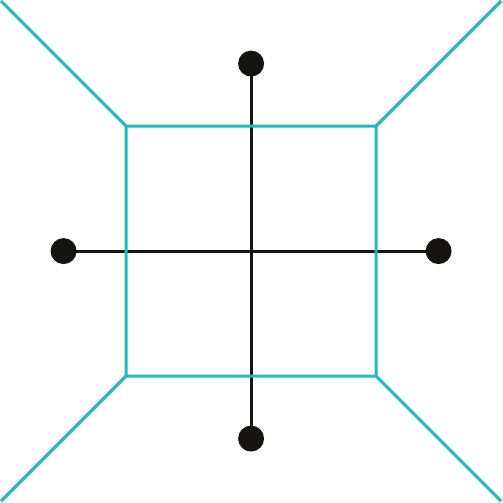 
  \caption{One-loop integrals contributing to $F^{(1)}$ with corresponding numerators.}\label{Fig:oneloop}
\end{figure}

\begin{figure}[h!]
  \centering
  \def\svgwidth{2cm} 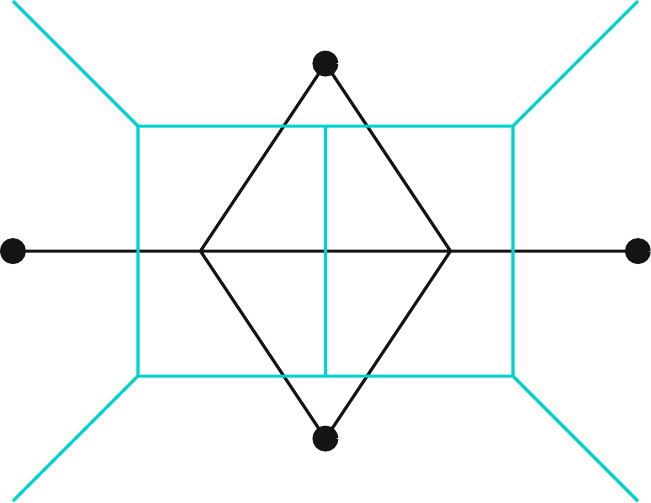  \qquad \qquad
  \def\svgwidth{2cm} 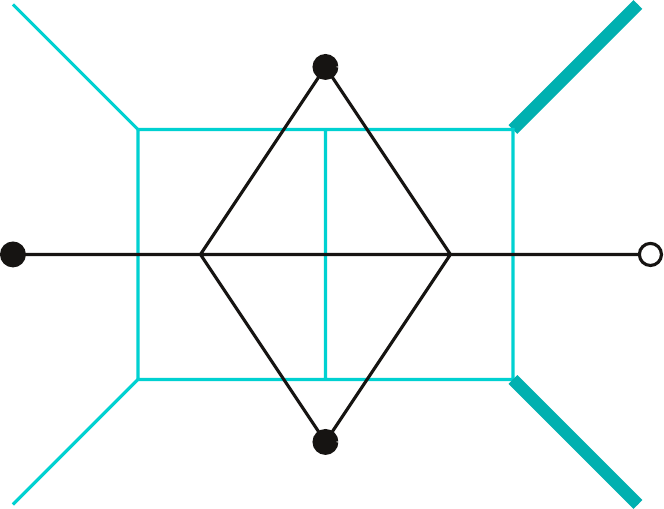  \qquad \qquad
  \def\svgwidth{2cm} 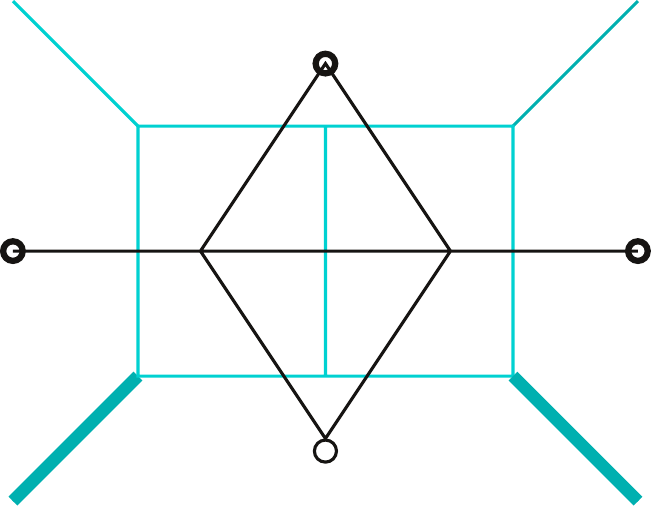  \qquad \qquad
  \def\svgwidth{2cm} 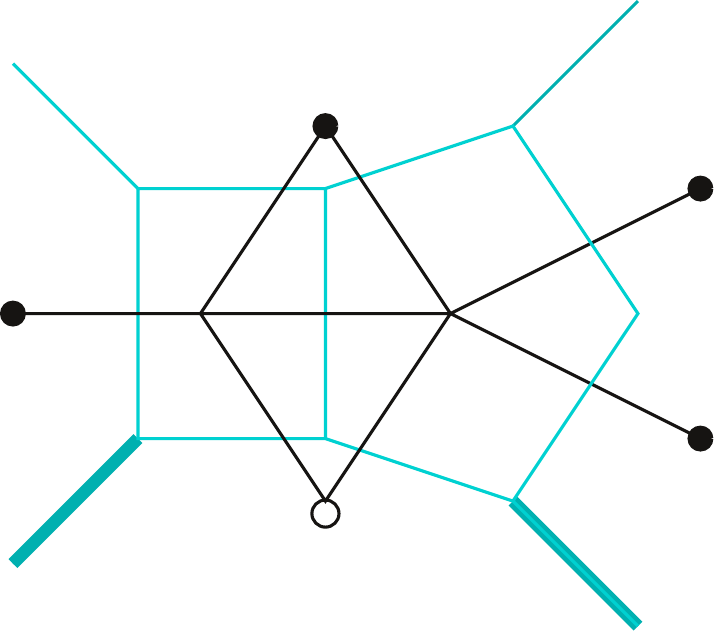 
  \def\svgwidth{4cm} 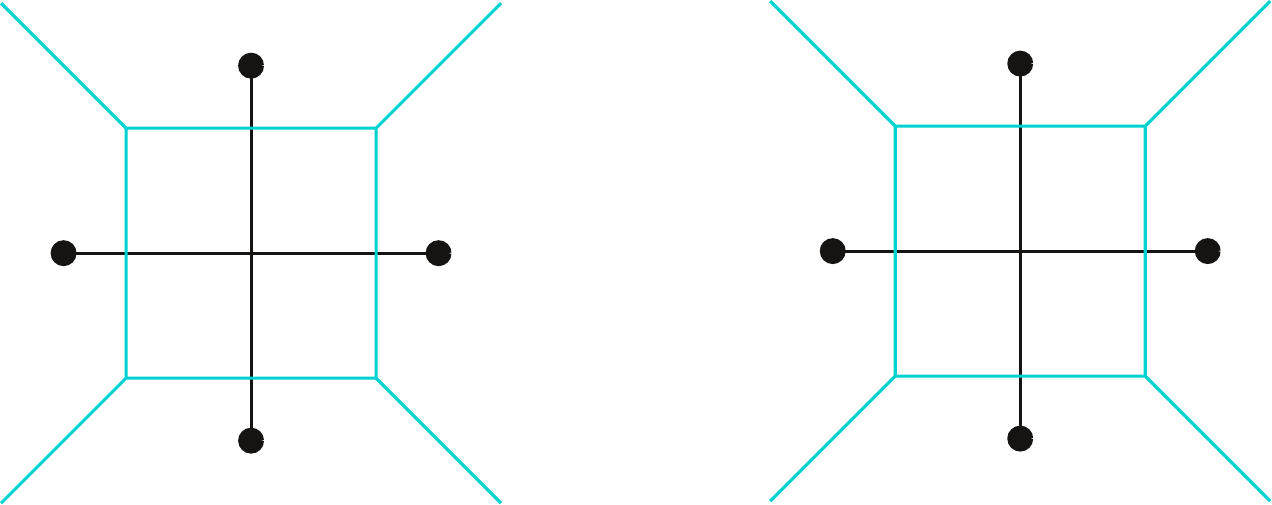 \qquad  \quad
  \def\svgwidth{4cm} 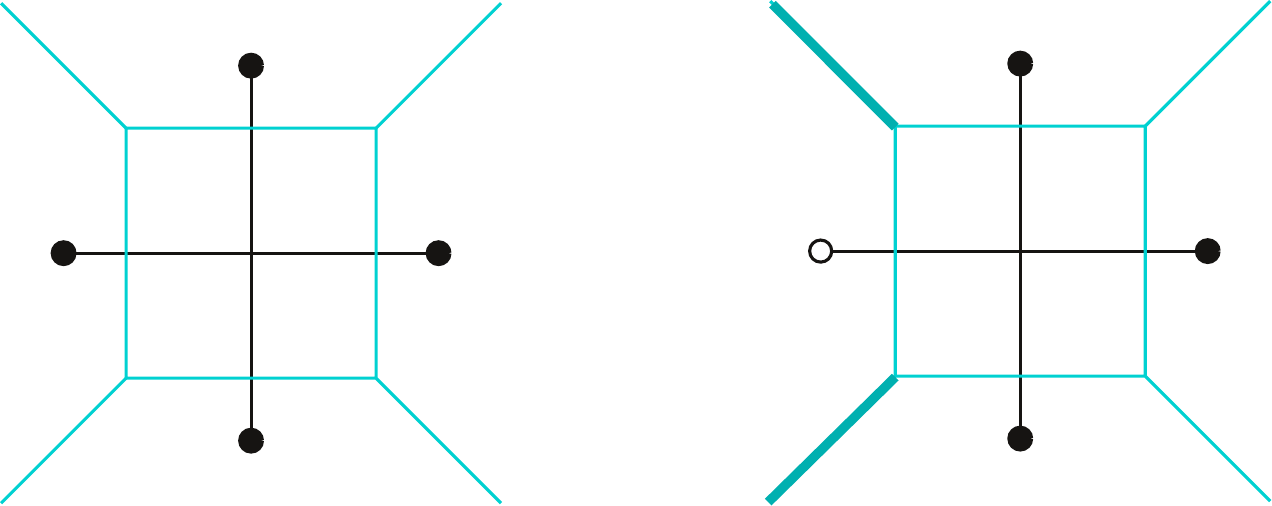 \qquad  \quad
  \def\svgwidth{4cm} 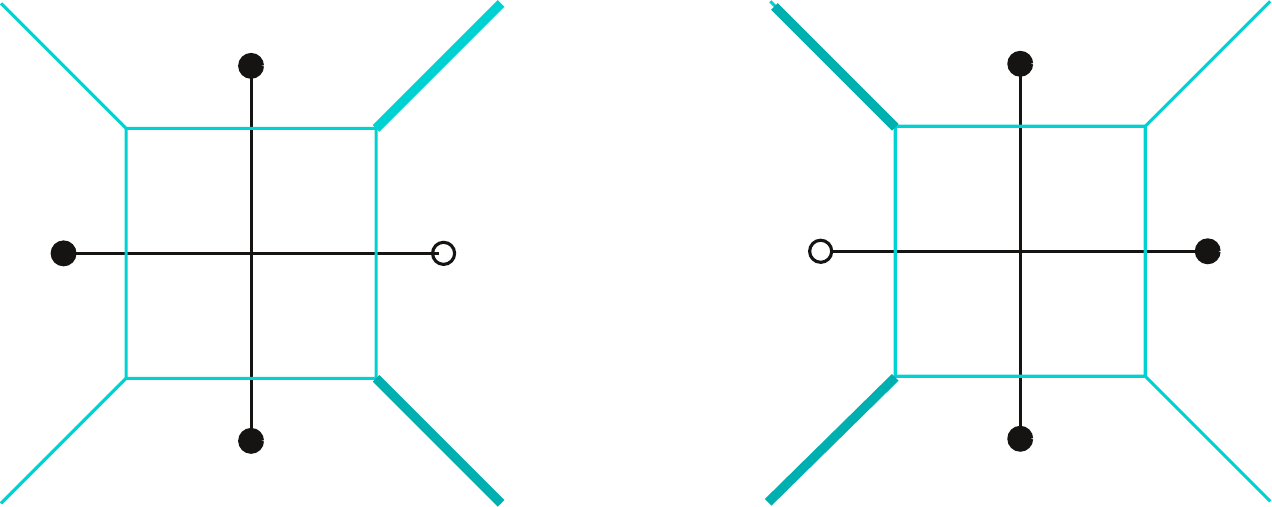 
  \caption{Two-loop integrals contributing to $F^{(2)}$ with corresponding numerators.}\label{Fig:twoloop}
\end{figure}

The one- and two-loop integrals are shown in Fig.~\ref{Fig:oneloop} and Fig.~\ref{Fig:twoloop},
respectively.
In dimensional regularization with $D=4-2\eps$ and $\eps<0$, they are defined by
\begin{align}
F_{1235} = \int \frac{d^{D}x_{6}}{i \pi^{D/2}} \, \frac{x_{13}^2 x_{25}^2}{x_{16}^2 x_{26}^2 x_{36}^2 x_{56}^2} \,,
\end{align}
and
\begin{align}
I_{1} = \int \frac{d^{D}x_{6} d^{D}x_{7}}{(i \pi^{D/2})^2} \, \frac{x_{13}^2 x_{24}^4}{x_{46}^2 x_{16}^2 x_{26}^2  x_{67}^2 x_{27}^2 x_{37}^2 x_{47}^2 } \,,
\end{align}
and similarly for the remaining integrals.

Although the individual integrals are divergent, the final answer for $F$ should be finite, as discussed in the introduction. 
We have performed
the calculation both using dimensional regularization, as well as in a mass regularization setup \cite{Alday:2009zm}. 
We found that the calculation was simpler in the massive regularization, as expected based on previous
experience with similar integrals. Below we outline the steps that allowed us to find an analytic answer
in the massive regularization. We verified numerically that the calculation in dimensional regularization gives the same finite answer. 

\subsection{Description of calculation}

The method we used for the calculation is standard and straightforward, 
and we only briefly mention the main steps. 
We first wrote down Mellin-Barnes representations for all integrals 
(see {\it e.g.} appendix A for an example or  \cite{Dixon:2011nj}, where a similar computation was done). 
We did this by introducing Mellin-Barnes representations one loop at a time. 
This has the advantage that it is straightforward to automate and gives
relatively compact answers.

Having obtained an expression in terms of (multiple) Mellin-Barnes integrals, we proceeded to extract the
divergences as $\eps \to 0$ in dimensional regularization, or $m^2 \to 0$ in the massive regularization, respectively.

In order to simplify the calculation, we used the fact that the answer is a conformally invariant function that depends
on $x_{i}^{\mu}$ through the variable $x$ only. Performing scaling limits such as e.g. $x_{i} \to \infty$ leaves this
function invariant, but it does simplify individual integrals. In taking several limits, we were able to simplify the
expression to a point where one could verify the absence of divergences in $F^{(2)}(x)$ analytically.

We arrived at a representation for the finite part of $F^{(2)}(x)$ in terms of a number of one-fold Mellin-Barnes integrals, and one two-fold one. 
The only two-fold Mellin-Barnes integral we encountered is
\begin{align}\label{equation:twofold-MBintegral}
& I(x)= \int \frac{dz_1 dz_2}{(2 \pi i)^2} \, x^{-1-z_1} \Gamma^2(-z_1) \Gamma^2(1 + z_1)\Gamma(-z_2) \Gamma(1 + z_2)\,\nonumber \\
& \quad\quad\quad\quad\quad\quad\quad \times \Gamma(1+z_1-z_2) \Gamma(-1-z_1+z_2)(\Psi_{0}( z_2)+\gamma_E)\,,
\end{align}
  with $-1<Re(z_1) <Re(z_2)<0$, and $\Psi_{0}$ is the polygamma function $\Psi_{n} = \partial^{(n+1)}_{z} \log \Gamma(z)$.

One can also reduce this integral to a one-fold one, as we explain presently.  
First, one  writes the polygamma functions as a derivative of a $\Gamma$ function w.r.t. an auxiliary parameter, 
\begin{align}
\Gamma(1+z) \Psi_{0}(z) =\lim_{\delta_1  \to 0}\frac{\partial}{\partial {\delta_1}} \Gamma(1+z+\delta_1) -\Gamma(z) \,.
\end{align}
Then one can see that the $z_{2}$ integration can be carried out using the first Barnes lemma. 
When doing this, it is useful to introduce another auxiliary parameter $\delta_2$ in order to separate the left and right poles of $ \Gamma( 1 + z_1 - z_2) \Gamma(-1 - z_1 + z_2)  \longrightarrow  \Gamma( 1 + z_1 - z_2 +\delta_2) \Gamma(-1 - z_1 + z_2 + \delta_2) $, and send $\delta_2 \to 0$ afterwards. (And similarly for the term with $\Gamma(-z_2)\Gamma(z_2) \longrightarrow \Gamma(-z_2 +\delta_2 ) \Gamma(z_2 + \delta_2) $.) 
In this way, we obtain a one-fold Mellin Barnes representation for $F^{(2)}$. 
At this stage, one can write the answer in terms of a series expansion that can
be resummed. We give an explicit example in Appendix A.

In this way, we arrived at our final result for $F^{(2)}$, which is given in eq.~(\ref{Ftwoloop}).

\section{Twistorial representation}

In the previous section we have presented the integrals that give us the observable under consideration up to second order and explained how to evaluate them using Mellin-Barnes methods. However, the set of Feynman integrals we used in eq.~(\ref{Fintegrals}) has a drawback. Although the final answer is finite, each individual Feynman integral has divergences. Only when we sum all of the integrals together the dependence on the regulator drops out. 
On the other hand, we expect that a good choice of master integrals can significantly simplify calculations involving Feynman integrals. For example, introducing infrared finite master integrals into the set of basis of Feynman integrals, has both conceptional and practical advantages, see~\cite{ArkaniHamed:2010gh,Drummond:2010mb,Drummond:2010cz}.
For example, one may hope that it will be easier in such representations to find powerful differential equations \cite{Drummond:2010cz}.
We give an example of this at the one-loop order.
Recall that $F$ is finite in four dimensions. We will therefore perform all manipulations in this section
in four dimensions. When individual IR divergent parts of the answer are evaluated, it should go
without saying that these should be regulated in a consistent way, in particular when using numerator
identities.

\subsection{One loop}
At one loop, one can use numerator identities described in~\cite{Drummond:2010mb} to express the one loop contribution to $F$ in terms of a finite pentagon integral. This can be done directly in the dual space, but it is more convenient to switch to twistor space.
Here we will only present the necessary conventions and refer the reader to~\cite{Hodges:2009hk,Mason:2009qx,ArkaniHamed:2010gh} for a more complete discussion. For any point $x_i$ in the dual space we associate a line in twistor space e.g. to $x_2$ we associate a line denoted by $(12)$ and spanned by two points in twistor space $Z_1$ and $Z_2$ , similarly to points $x_5$, $x_6$, $x_7$ we associate lines $(AB)$, $(CD)$, $(EF)$. The points $x_{1},..., x_{4}$ define a null polygon, and so the corresponding lines in twistor space intersect. Therefore, it is natural to associate the point $x_i$ for $i=1,...,4$ with a line $(i\!-\!1\,\: i)$, where the number $i\!-\!1$ is defined modulo 4. Using the incidence relations that relate the momentum twistor space to dual space we can find that e.g.
\begin{align} \label{equ:to_twistors}
  x_{56}^2 \equiv (x_5-x_6)^2 = \dfrac{\langle ABCD\rangle}{\langle AB\rangle \langle CD\rangle},
\end{align}
where $\langle ABCD\rangle = \sum \epsilon_{ijkl} Z^i_A Z^j_B Z^k_C Z^l_D$ and $\langle AB\rangle=\langle AB I_\infty\rangle$, where $I_\infty$ is the infinity twistor. In a similar way we can rewrite our integrand in terms of twistor brackets. Moreover, as all integrals we deal with are conformally invariant, the dependence on $I_\infty$ drops out and we are left only with the 4-brackets. The resulting expression for the integrand can be subsequently simplified by the following identity
\begin{align}\label{equ:oneloop_identity}
  \langle AB13\rangle\langle CD24\rangle+\langle AB24\rangle\langle CD13\rangle =& \langle12AB\rangle\langle34CD\rangle +\langle34AB\rangle\langle12CD\rangle \nonumber \\ & \hspace{-2 cm} -\langle23AB\rangle\langle41CD\rangle-\langle41AB\rangle\langle23CD\rangle-\langle1234\rangle\langle ABCD\rangle \,.
\end{align}
Eq. (\ref{equ:oneloop_identity}) is a special case of an identity described in~\cite{Drummond:2010mb}. Following the described steps, we end up with a very simple expression for the one loop integrand
\begin{align}\label{equ:oneloop_pentagon}
  F^{(1)} =& \, \dfrac{1}{4} \, (F_{1235} + F_{4125} + F_{3415} + F_{2345} -F_{1234})  \nonumber \\
	  =&    \, - \dfrac{1}{4} \, \int \dfrac{d^4 Z_{CD}}{i \pi^2} \dfrac{ (\langle AB13\rangle\langle CD24\rangle+\langle AB24\rangle\langle CD13\rangle) \langle 1234\rangle}{\langle CD12\rangle\langle CD23\rangle\langle CD34\rangle\langle CD41\rangle\langle CDAB\rangle}. 
\end{align}
\begin{figure}[h!]
 \centering
  \begin{fmffile}{oneloop}
  \begin{fmfchar*}(25,25)
    \fmfsurround{i3,d2,i2,d1,i1,d5,i5,d4,i4,d3}
    \fmf{plain}{i1,v1}
    \fmf{plain}{i2,v2}
    \fmf{plain}{i3,v3}
    \fmf{plain}{i4,v4}
    \fmf{plain}{i5,v5}
    \fmf{dashes, tension=0.1}{v2,v4}
    \fmf{plain, tension=0.2}{v1,v2,v3,v4,v5,v1}
    \fmfv{label=$1$,label.dist=-3}{d1}
    \fmfv{label=$2$,label.dist=-4}{d2}
    \fmfv{label=$3$,label.dist=-4}{d3}
    \fmfv{label=$4$,label.dist=-4}{d4}
    \fmfv{label=$5$,label.dist=-4}{d5}
  \end{fmfchar*}
  \end{fmffile}
 \caption{One-loop chiral integral contributing to $F^{(1)}$.}
 \label{fig:one-loop-chiral}
\end{figure}
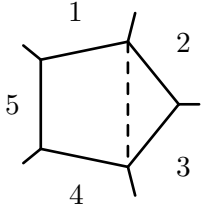
The chiral pentagon integral that we obtain in eq.~(\ref{equ:oneloop_pentagon}) is presented in figure~\ref{fig:one-loop-chiral}.
For generic $x_{14}^2$ it is given by
\begin{align}
  \Psi^{(1)}(u,v) = \log u \log v + {\rm Li}_{2}(1-u) + {\rm Li}_{2}(1-v) -\zeta_2
\end{align}
with $u=x_{15}^2 x_{24} / (x_{25}^2 x_{14}^2)\,, v=x_{45}^2 x_{13}^2/(x_{35}^2 x_{14}^2)$. 
In the limit $x_{14}^2 \to 0$, both $u$ and $v$ diverge, with $v/u = x$ fixed. We find that
\begin{align}\label{resultF1}
  F^{(1)} =& \, - \dfrac{1}{2} \, \lim_{x_{14}^2 \to 0}\Psi^{(1)}(u,v) = \frac{1}{4} \left[ \log^2 x + \pi^2 \right] \,,
\end{align}
in perfect agreement with the one-loop result~(\ref{Foneloop}).
Of course, this integral is so simple that it can be evaluated by many methods, e.g. using Feynman parameters, see e.g. \cite{Drummond:2010mb,Alday:2010jz}.
A much more elegant way way of computing it is based on differential equations \cite{Drummond:2010cz}.
Here we wish to mention that the latter are compatible with the limit  $x_{14}^2 = 0$.
In other words, the differential equations of \cite{Drummond:2010cz} directly apply to that case,
and one obtains
\begin{align}
x \partial_x x\partial_x   F^{(1)}(x)  = \frac{1}{2} \,.
\end{align}
This, together with the boundary condition $F^{(1)}(x\to -1) = 0$, which follows from inspection of the twistor numerator, 
leads to the result of eq. (\ref{resultF1}).
It would be interesting to compute the integrals appearing at higher loop orders in a similar way.

\subsection{Two loops}
At two loops, we would like to find an identity, similar to eq.~(\ref{equ:oneloop_identity}), that gives a chiral representation of the integrand, similar to eq.~(\ref{equ:oneloop_pentagon}). In order to find such an identity, we proposed an ansatz for the integrand in terms of a set chiral integrals, that could potentially represent our integrand at two loops, with arbitrary coefficients. In order to fix the coefficients we performed quadruple cuts. Having determined all coefficients, we then checked the equality of the two integrands analytically, by expanding out the twistor four-brackets.
The representation we found is
\begin{align}\label{equ:twoloop_chiral}
 F^{(2)} = \dfrac{1}{64} \, \sum_{8\; {\rm perm}}\, \big(-I_a + 8 I_{b} -4 I_c +8 I_{d}+I_{e}\big),
\end{align}
where the $8$ permutations refer to $4$ cyclic permutations of the points $Z_{1},Z_{2},Z_{3},Z_{4}$, and swaps $Z_{1} \leftrightarrow Z_{4}, Z_{2} \leftrightarrow Z_{3}$.
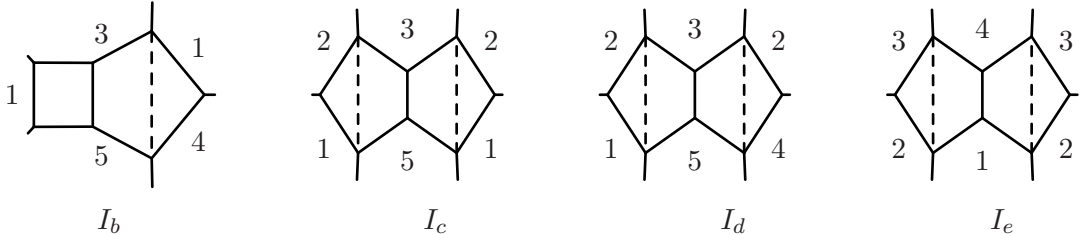
\begin{figure}[h!]
\vspace*{20pt}
  \centering
  \begin{fmffile}{twoloop}
  \begin{fmfchar*}(25,25)
    \fmfleft{d1,d2,d3,l1,d4,d5,d6,l2,d7,d8,d9}
    \fmftop{d11,d12,t1,d13}
    \fmfbottom{d21,d22,b1,d23}
    \fmfright{d31,d32,r1,d33,d34}
    \fmf{plain}{r1,v1}
    \fmf{plain}{t1,v2}
    \fmf{plain}{l2,v4}
    \fmf{plain}{l1,v5}
    \fmf{plain}{b1,v7}
    \fmf{plain, tension=0.1}{v1,v2,v3,v4,v5,v6,v7,v1}
    \fmf{dashes, tension=0.15}{v2,v7}
    \fmf{plain, tension=0.05}{v3,v6}
    \fmfv{label=$1$,label.dist=2}{d5}
    \fmfv{label=$3$,label.dist=-16}{d12}
    \fmfv{label=$4$,label.dist=-8}{d32}
    \fmfv{label=$1$,label.dist=-8}{d33}
    \fmfv{label=$5$,label.dist=-16}{d22}
  \end{fmfchar*}
  \qquad \quad
  \begin{fmfchar*}(25,25)
    \fmfsurround{i2,d2,i3,d3,i4,d4,i5,d5,i6,d6,i1,d1}
    \fmf{plain}{i1,v2}
    \fmf{plain}{i2,v3}
    \fmf{plain}{i3,v4}
    \fmf{plain}{i4,v6}
    \fmf{plain}{i5,v7}
    \fmf{plain}{i6,v8}
    \fmf{plain, tension=0.1} {v1,v2,v3,v4,v5,v6,v7,v8,v1}
    \fmf{dashes, tension=0.15}{v2,v4}
    \fmf{dashes, tension=0.15}{v6,v8}
    \fmf{plain, tension=0.15}{v1,v5}
    \fmfv{label=$1$,label.dist=-4}{d1}
    \fmfv{label=$2$,label.dist=-4}{d2}
    \fmfv{label=$3$,label.dist=-14}{d3}
    \fmfv{label=$2$,label.dist=-4}{d4}
    \fmfv{label=$1$,label.dist=-4}{d5}
    \fmfv{label=$5$,label.dist=-14}{d6}
  \end{fmfchar*}
  \qquad \quad
  \begin{fmfchar*}(25,25)
    \fmfsurround{i2,d2,i3,d3,i4,d4,i5,d5,i6,d6,i1,d1}
    \fmf{plain}{i1,v2}
    \fmf{plain}{i2,v3}
    \fmf{plain}{i3,v4}
    \fmf{plain}{i4,v6}
    \fmf{plain}{i5,v7}
    \fmf{plain}{i6,v8}
    \fmf{plain, tension=0.1} {v1,v2,v3,v4,v5,v6,v7,v8,v1}
    \fmf{dashes, tension=0.15}{v2,v4}
    \fmf{dashes, tension=0.15}{v6,v8}
    \fmf{plain, tension=0.15}{v1,v5}
    \fmfv{label=$4$,label.dist=-4}{d1}
    \fmfv{label=$2$,label.dist=-4}{d2}
    \fmfv{label=$3$,label.dist=-14}{d3}
    \fmfv{label=$2$,label.dist=-4}{d4}
    \fmfv{label=$1$,label.dist=-4}{d5}
    \fmfv{label=$5$,label.dist=-14}{d6}
  \end{fmfchar*}
  \qquad \quad
  \begin{fmfchar*}(25,25)
    \fmfsurround{i2,d2,i3,d3,i4,d4,i5,d5,i6,d6,i1,d1}
    \fmf{plain}{i1,v2}
    \fmf{plain}{i2,v3}
    \fmf{plain}{i3,v4}
    \fmf{plain}{i4,v6}
    \fmf{plain}{i5,v7}
    \fmf{plain}{i6,v8}
    \fmf{plain, tension=0.1} {v1,v2,v3,v4,v5,v6,v7,v8,v1}
    \fmf{dashes, tension=0.15}{v2,v4}
    \fmf{dashes, tension=0.15}{v6,v8}
    \fmf{plain, tension=0.15}{v1,v5}
    \fmfv{label=$2$,label.dist=-4}{d1}
    \fmfv{label=$3$,label.dist=-4}{d2}
    \fmfv{label=$4$,label.dist=-14}{d3}
    \fmfv{label=$3$,label.dist=-4}{d4}
    \fmfv{label=$2$,label.dist=-4}{d5}
    \fmfv{label=$1$,label.dist=-14}{d6}
  \end{fmfchar*}
  \end{fmffile}
  \linebreak \qquad\qquad\qquad $I_{b}$ \qquad\qquad\qquad\qquad\qquad $I_{c}$ \qquad\qquad\qquad\qquad\quad $I_{d}$ \qquad\qquad\qquad\qquad $I_{e}$ 
  \caption{Twistorial representation of two-loop integrals contributing to $F^{(2)}$.}\label{fig:twoloop-chiral}
\end{figure}
Here, I's are the chiral twistor integrals, presented in figure~\ref{fig:twoloop-chiral} and are defined as follows
{\small{\begin{align}
  I_{a}&=\,\int\dfrac{d^4 Z_{CD}}{i \pi^2}\dfrac{d^4 Z_{EF}}{i \pi^2}\:\dfrac{\langle AB13\rangle\langle CD24\rangle+\langle AB24\rangle\langle CD13\rangle}{\langle CD12\rangle\langle CD23\rangle\langle CD34\rangle\langle CD41\rangle\langle CDAB\rangle}\times \Big((CD)\leftrightarrow(EF)\Big)\nonumber\\
  I_{b}&=\,\int\dfrac{d^4 Z_{EF}}{i \pi^2}\dfrac{d^4 Z_{CD}}{i \pi^2}\:\dfrac{\langle AB34\rangle\langle1234\rangle\big(\langle EF13\rangle\langle AB24\rangle+\langle AB13\rangle\langle EF24\rangle\big)}{\langle CDAB\rangle\langle CD41\rangle\langle CD23\rangle\langle CDEF\rangle\langle EF23\rangle\langle EF34\rangle\langle EF41\rangle\langle EFAB\rangle}\nonumber\\
  I_{c}&=\,\int\dfrac{d^4 Z_{EF}}{i \pi^2}\dfrac{d^4 Z_{CD}}{i \pi^2}\:\dfrac{\langle1234\rangle\langle12AB\rangle^2\big(\langle EF13\rangle\langle CD24\rangle+\langle EF24\rangle\langle CD13\rangle\big)}{\langle CDAB\rangle\langle CD41\rangle\langle CD12\rangle\langle CD23\rangle\langle CDEF\rangle\langle EF41\rangle\langle EF12\rangle\langle EF23\rangle\langle EFAB\rangle}\nonumber\\
  I_{d}&=\,\int\dfrac{d^4 Z_{EF}}{i \pi^2}\dfrac{d^4 Z_{CD}}{i \pi^2}\:\dfrac{\langle1234\rangle\langle12AB\rangle\langle23AB\rangle\big(\langle EF13\rangle\langle CD24\rangle+\langle EF24\rangle\langle CD13\rangle\big)}{\langle CDAB\rangle\langle CD41\rangle\langle CD12\rangle\langle CD23\rangle\langle CDEF\rangle\langle EF34\rangle\langle EF12\rangle\langle EF23\rangle\langle EFAB\rangle}\nonumber\\
  I_{e}&=\,\int\dfrac{d^4 Z_{CD}}{i \pi^2}\dfrac{d^4 Z_{EF}}{i \pi^2}\:\dfrac{\langle1234\rangle^3\big(\langle CD13\rangle\langle EF24\rangle+\langle CD24\rangle\langle EF13\rangle\big)}{\langle CD12\rangle\langle CD23\rangle\langle CD34\rangle\langle CD41\rangle\langle CDEF\rangle\langle EF12\rangle\langle EF23\rangle\langle EF34\rangle\langle EF41\rangle}.\nonumber
\end{align}}}

\subsection{Checking finiteness}

Let us now discuss the finiteness properties of the new representation presented in eq.~(\ref{equ:twoloop_chiral}).
$I_a$ is the square of the one loop pentagon integral in figure~\ref{fig:one-loop-chiral}, which is finite, and it is easy to see that $I_b$ is also finite.
Moreover, one can see that none of the integrals $I_c$, $I_{d}$, $I_{e}$ has one-loop subdivergences.
 However, they do separately have overall double logarithmic infrared singularities. This is very similar to
 the behavior of the logarithm of the four-particle amplitude discussed in \cite{Drummond:2010mb}.
 Here, the remaining divergence cancels for the particular combination ${\sum_{8\; {\rm perm}}(-4 I_c +8 I_{d}+I_{e})}$, making the final answer finite. 
 Twistor space also makes it easy to demonstrate the above statements about
IR properties, and we devote the remainder of this subsection to show this.

In order to investigate the divergences of the integrals in figure~\ref{fig:twoloop-chiral} we adopt the parametrization used in \cite{Bourjaily:2011hi}. The dual conformal integrals discussed in this paper have only infrared divergences. These arise when the loop momentum becomes collinear with the momentum of an external massless particle.
These limits can be conveniently parametrized in twistor space. For example, consider the integration variable $x_6$, which corresponds to $Z_c$ and $Z_d$. One of the collinear limits is described by $Z_c$ tending to $Z_2$, while $Z_d$ tends to a generic point on the hyperplane spanned by $Z_1$, $Z_2$, $Z_3$,
\begin{align}
 Z_c &\rightarrow Z_2 + \cO(\epsilon) \\
 Z_d &\rightarrow \alpha_1 Z_1 +\alpha_2 Z_2 + \alpha_3 Z_3 + \cO(\epsilon) \,.
\end{align} 
For the present purpose, we also need to parametrize the $\cO(\epsilon)$ terms. In doing so, we must
make sure that the parametrization is generic enough.
That is, we do not impose any additional constrains, which can make some factors vanish faster than in the generic case. On the other hand, we need to restrict the integration variable $x_6$ to be real. This corresponds to considering an integration bitwistor $Y^{I J} = Z_c^{[I} Z_d^{J]}$ of the form
\begin{align}\label{equ:collinear_limit_bitwistor}
 Y = \alpha_1\, (12) + \alpha_3\, (23) + \epsilon \: \Big(\alpha_4\, (12) + \alpha_5\, (23) + \alpha_6\, (34) + \alpha_7\, (41) \Big)  + \cO(\epsilon^2)\,.
\end{align}
Here the bracket $(ij)$ denotes a line in twistor space spanned by $Z_i$ and $Z_j$.
Note that $Y$ does not contain ``non-local''  terms like $(13)$ or $(24)$. 
Moreover, due to the fact that we are dealing with an integrand that is explicitly symmetric under cyclic permutations of $Z_1, ..., Z_4$ we only have to consider the collinear limit presented in eq. (\ref{equ:collinear_limit_bitwistor}). Imposing that limit we confirmed that $I_b$ vanishes while  ${I_{c},\, I_{d},\, I_{e}}$ diverge for generic $\alpha$'s as ${I_{c},\, I_{d},\, I_{e} \sim \epsilon^{-1}}$. However, the considered combination of integrals behaves as ${\sum_{8\; {\rm perm}} (-4 I_c +8 I_d + I_e) \sim \epsilon ^0}$. Therefore, it is finite in all collinear limits, even though $I_{c},\, I_{d},\, I_{e}$ separately suffer from logarithmic divergences\footnote{It is very interesting to note that the combination $-4 I_c +8 I_{d}+I_{e}$ is finite if and only if we sum over the four cyclic permutations.}.

\subsection{Observation on two loop result}

We would like to finish this section on chiral representation of integrands by making an interesting observation that relates the two loop result found in section \ref{section:two-loop_calculation} to finite integrals recently computed in the literature.
Recall that sending $x_{5}$ to infinity via a conformal transformation our integrals become
(in general non-planar) integrals for the scattering of four massless particles.
The planar master integrals for such a process are all known, see \cite{Gehrmann:2000xj} and references therein.
 Recently \cite{CaronHuot:2012ab} computed two finite two-loop master integrals, called $I_{++}$ and $I_{+-}$ with double-box topology using twistor methods.
It is interesting to note that these two functions together with their transforms $x \to 1/x$ are related to our two-loop result in the following way,
\begin{align}
  F^{(2)}(x) + 2 \zeta(2)  F^{(1)}(x) +  \big( F^{(1)}(x) \big)^2  = \left[ -\frac{1}{4} I_{++}(x)  + \frac{1}{4} I_{+-}(x)  \right]  + \left[ x \leftrightarrow \frac{1}{x} \right] + 3 \zeta_3\,.
\end{align}
As is well-known, $3 \zeta_{3}$ could also be written as a finite two-loop integral, see for example 
section 3.5. of \cite{Smirnov:2006ry}.
It would be natural to absorb this into the definition of $I_{++}$ and $I_{+-}$.
We suspect that one can find a connection between the integrals for our observable from section \ref{section:two-loop_calculation} and the 4-point integrals computed in \cite{CaronHuot:2012ab}, by sending $x_{5}$ to infinity via a conformal transformation. However, further work needs to be done to clarify this point.

\section{Relation to the light-like cusp anomalous dimension}

The Lagrangian insertion procedure would naively imply
\begin{align}\label{insertion_naive}
\int \frac{d^{4}x_{5}}{i \pi^{2}} \, \frac{F(x)}{x_{15}^2 x_{25}^2 x_{35}^2 x_{45}^2} \stackrel{naively}{=} \lambda \frac{\partial}{\partial\lambda} \log \langle W_4 \rangle \,.
\end{align}
We wrote ``naively'' because both sides of eq. (\ref{insertion_naive}) diverge double logarithmically, 
due to soft-collinear divergences.
A valid equation can be written down within a given regularization.
However, $F$ is defined in the limit where the regulator tends to zero. 
Therefore,
we can at most hope that re-instating a regulator
in eq.~(\ref{insertion_naive}) will allow us to compare the leading divergence of its l.h.s. and r.h.s.. 

As we will see, this can be successfully done using a massive or dimensional regulator.
See also for a discussion of regulator-independent quantities \cite{Henn:2011by}, calculated
in massive and dimensional regularization. 

\subsection{Massive regularization}

Consider the four-particle scattering amplitude dual to the four-cusp Wilson loop,
defined with a massive regulator, $M_{4}$ \cite{Alday:2009zm}.
Based on the structure of infrared divergences of the latter, we expect the following equation to hold,
\begin{align}\label{integral_insertion1}
I_{m^2}[F] := \int \frac{d^{4}x_{5}}{i \pi^{2}} \, \frac{F(x)}{\prod_{i=1}^{4} (x_{i5}^2 + m^2) } = 
\lambda \frac{\partial}{\partial\lambda} \log \langle M_4 \rangle + \cO(\log m^2 )\,.
\end{align}
Here we only need the leading infrared divergences of $ \log \langle M_4 \rangle $, which are given by 
(see e.g. \cite{Henn:2011by,Henn:2010bk})
\begin{align}\label{div-massive}
 \log \langle M_4 \rangle  = -  \frac{1}{2}  \log^2 m^2 \, \Gamma_{\rm cusp}  + \cO(\log m^2 ) \,,
\end{align}
with the cusp anomalous dimension
\begin{align}\label{gamma_3loop}
\Gamma_{\rm cusp} =&  \sum_{L=1}^{\infty}  \left(\frac{\lambda}{8 \pi^2}\right)^L \, \Gamma^{(L)}_{\rm cusp} \,,\\ 
 =&
 2 \left(\frac{\lambda}{8 \pi^2}\right) 
- 2 \zeta_2   \left(\frac{\lambda}{8 \pi^2}\right)^2 
+ 11 \zeta_4  \left(\frac{\lambda}{8 \pi^2}\right)^3 + \cO(\lambda^4)\,.
\end{align}
One might worry that eq. (\ref{integral_insertion1}) does not make sense, since we did not keep the dependence on $m^2$ 
in $F$. However, one can argue that this additional dependence will not affect the leading divergence.
We caution the reader that the same argument is slightly more subtle in dimensional regularization, as discussed in
the following subsection.

We wish to verify the above relation (\ref{integral_insertion1}) using our result for $F^{(2)}$.
In order to do this, it is convenient to compute the auxiliary integral
\begin{align}\label{power_int}
\int \frac{d^{4}x_{5}}{i \pi^{2}} \, \frac{x^p}{\prod_{i=1}^{4} (x_{i5}^2 + m^2) } = 2 \log^2 m^2 \,  \frac{\sin(\pi p)}{\pi p} + \cO(  \log  m^2 )\,.
\end{align}
We will give the derivation of an analogous formula in dimensional regularization in appendix A. 
This formula is derived for $|p|<1$, but can be extended to other values of $p$ by analytic continuation.
Due to the $x \to 1/x$ symmetry of $F$ we can assume $0<x<1$ without loss of generality. 
Then, we write $F(x)$ as a series in $x$ around $0$ and use eq.~(\ref{power_int}) to 
perform the integration.

In doing so, one sees that only constants and logarithmically enhanced terms in $F$ contribute to the cusp anomalous dimension.
Note however that we do need to keep terms like $\log^k(x) x^n$, to all orders in $n$.\footnote{An instructive example of
this is the function $\log(x/(1+x)) \log(1+x)$, which could have appeared at two loops. Although it vanishes as $x \to 0$ (and as $x\to \infty$, due to the inversion symmetry), it would give a contribution of $\zeta_2$ to the cusp anomalous dimension.}
The one- and two-loop calculations are elementary. Technical details of the three-loop calculation are given in the appendix.
To three loops, we find
\begin{align}
I_{m^2}[F]  =  \log^2 m^2 \, \left[ - \left( \frac{\lambda}{8 \pi^2}\right) + \frac{1}{3} \pi^2 \left( \frac{\lambda}{8 \pi^2}\right)^2 + \frac{33}{2} \zeta_4 \left( \frac{\lambda}{8 \pi^2}\right)^3 + \cO(\lambda^4) \right] + \cO(\log m^2 )\,.
\end{align}
Taking into account eqs. (\ref{div-massive}) and (\ref{gamma_3loop}), we see that 
this is in perfect agreement with eq. (\ref{integral_insertion1}). 

\subsection{Dimensional regularization}
The same calculation can also be done in dimensional regularization.
In order to get the correct result however, we have to be careful.
The calculation of $F$ has been done for $D=4$, and we should
really do the whole calculation using $D=4-2 \eps$.
Can we still recover the leading $1/\eps^2$ term correctly?
The answer turns out to be yes, but we need to go just a little
beyond the $\eps=0$ approximation in the calculation of $F$.
In fact, on dimensional grounds $F^{(L-1)}$ must have dimension $(L-1)\eps$.
It is important to take this into account.
As far as the leading pole is concerned, 
one can see that this effectively amounts to
multiplying the naive answer at $L$ loops by a factor of $1/L^2$. 
Hence we expect
\begin{align}
\int \frac{d^{D}x_{5}}{i \pi^{D/2}} \, \frac{F^{(L-1)}(x)}{x_{15}^2 x_{25}^2 x_{35}^2 x_{45}^2} 
= - \sum_{L=1}^{\infty} \left( \frac{\lambda}{8 \pi^2} \right)^L  8 \frac{L}{\eps^2} \Gamma^{(L)}_{\rm cusp} + \cO(\eps^{-1}) \,.
\end{align}
where the r.h.s. again follows from the known structure of divergences of light-like Wilson loops \cite{Korchemskaya:1992je},
where we have multiplied the usual $L$-loop contribution by $L^2$, as explained above.

Then, with the dimensional regularization version of eq.~(\ref{power_int}), derived in Appendix A,
\begin{align}\label{power_int_DR}
\int \frac{d^{4-2 \eps}x_{5}}{i \pi^{2}} \, \frac{x^p}{\prod_{i=1}^{4} x_{i5}^2  } = 4 \frac{1}{\eps^2} \,  \frac{\sin(\pi p)}{\pi p} + \cO(\eps^{-1}) \,,
\end{align}
we can reproduce the correct answer for the cusp anomalous dimension to three loops, see eq. (\ref{gamma_3loop}).

\subsection{Strong coupling}

In ref. \cite{Alday:2011ga} the following answer for $F(x)$ was found at strong coupling,
\begin{align}
F(x) =  \frac{x}{(1-x)^3} \, \left[ 2 (1-x) + (1+x) \log x \right] \frac{\sqrt{\lambda}}{4 \pi} + \ldots  \,, \quad \lambda \gg 1\,,
\end{align}
We also note the value of the cusp anomalous dimension at strong coupling, see ref. \cite{Gubser:2002tv,Frolov:2002av},
\begin{align}\label{gamma_strong}
\Gamma_{\rm cusp} = \frac{\sqrt{\lambda}}{2 \pi} + \ldots \,, \quad \lambda \gg 1\,.
\end{align}
Let us now verify the relation between $F(x)$ and $\Gamma_{\rm cusp}$, using eq.~(\ref{integral_insertion1}).

We could employ the Mellin transform of $F(x)$, but we find it easier just to use a series expansion near $x=0$.
Recall that in this approach the non-logarithmically enhanced terms in $F(x)$ do not play a role. (They are
needed however in order for $F(x)$ to be well-defined at $x \to 1$.)

There is a subtle point in this calculation, which concerns interchanging the expansion of $F(x)$ for small $x$ and
the space-time integration in eq. (\ref{power_int}). If one does this naively, one obtains sum of the type 
$\sum_{n \ge 0} (-1)^n$. A slightly more careful treatment, to be given presently, shows that this can be interpreted 
as $1/2 = 1/(1+1) = 1-1+1-1+ \ldots$.

Let us first compute the integral over the insertion for some generating functions,
for which the convergence of the series is clear, and then differentiate w.r.t. certain
parameters. In fact we will see that letting $x \to a x$ and using $a$ as such a 
parameter will be sufficient.
This will give a result valid for $a<1$, which we can extend to $a\to1$.
Indeed, we find that with
\begin{align}
g_1(x) =  -x \,, \qquad
g_2(x) = - \frac{x \log x}{1-x} \,, \qquad
g_2(x) = - \frac{x^2 \log x}{1-x} \,,
\end{align}
we can write
\begin{align}
 \frac{x}{(1-x)^3} \, \left[ -2 (1-x) - (1+x) \log x \right]  = \lim_{a \to 1} \left[ g_1(x) + \frac{\partial }{\partial a} g_{2}(a x) +  \frac{\partial^2}{\partial^2 a} g_{3}(a x) \right] \,. 
\end{align}
Moreover, we have, for $a<1$,
\begin{align}
I_{m^2}[ g_1(a x) ] =& 0 + \cO(\log m^2)  \,,\\
I_{m^2}[ g_2(a x) ] =& 2 \log^2 m^2 \, \log(1+a)+ \cO(\log m^2) \,,\\
I_{m^2}[ g_3(a x) ] =&2 \log^2 m^2 \, \left[  -a + \log(1+a) \right]+ \cO(\log m^2) \,,
\end{align}
so that we arrive at
\begin{align}
I_{m^2}\left[ \frac{x}{(1-x)^3} \, \left[ -2 (1-x) - (1+x) \log x \right] \right] = \frac{1}{2} \, \log^2 m^2 \,  + \cO(\log m^2) \,.
\end{align}
Comparing to eq. (\ref{integral_insertion1}) and (\ref{div-massive}), we find perfect agreement with the strong coupling value of the cusp anomalous dimension given in eq. (\ref{gamma_strong}).

\section{Summary and outlook}

In this paper we considered the correlation function of a local operator (the Lagrangian) with a four-sided null Wilson loop, in planar ${\cal N}=4$ SYM. This is an interesting quantity due to several reasons: it is finite; it interpolates between a scattering amplitude/Wilson loop and a correlation function and it is  a non-trivial function, not fixed by symmetries, of a single cross-ratio. Hence it is an ideal quantity to try to interpolate from weak to strong coupling. 

We computed analytically the two loop contribution to the above observable. The result has the expected degree of transcendentality and reproduces the correct value of the cusp anomalous dimension. Furthermore, we have given a twistorial representation for the result, which possesses several advantages. 

There are several open problems. It would be interesting to understand better the limit $x_{i5}^2 \to 0$ at loop level. 
At tree-level, or, what is the same, the level of the loop integrands, this is related to
a forward limit of a NMHV amplitude. It would be interesting if, despite regulator issues,
the corresponding limit at loop level was related to results for loop-level NMHV amplitudes.
The relevant six-point NMHV amplitudes are known analytically to the two-loop order \cite{Dixon:2011nj}. More generally, one would like to understand different OPE limits of our mixed correlator and understand which quantities/anomalous dimensions can be obtained from the answer presented in this paper. Finally, it would be extremely nice to guess a recursion relation/expression for this correlator, for instance by using twistor techniques along the lines of \cite{Adamo:2011cd}.

\section*{Acknowledgments}

We would like to thank N.~Arkani-Hamed, J.~Trnka, G.~Korchemsky, and especially
P. Heslop for interesting discussions.
J.M.H. was supported in part by the Department of Energy grant DE-FG02-90ER40542.
J.M.H. would like to thank the ECT* Trento for hospitality during the initial stage of this work.
The work of L.F.A. is partially supported by the ERC grant DUALITIESHEPTH.

\appendix

\section{Integration of $F(x)=x^p$ over the insertion point}

We would like to compute the leading divergent term of the following integral
\begin{displaymath}
 I_D(p)= s t \int \dfrac{d^D y}{i \pi^2} \frac{x^p}{\prod_{i=1}^4|y-x_i|^2}
\end{displaymath}
where $D=4-2\epsilon$ and recall that $x= ({x_{25}^2 x_{45}^2 x_{13}^2})/({x_{15}^2 x_{35}^2 x_{24}^2})$. 
This integral is finite in the range $-1 < p < 1$. Using Feynman parameters we obtain
\begin{displaymath}
 \frac{x^p}{\prod_{i=1}^4|y-x_i|^2}=\frac{6}{\Gamma^2(1+p)\Gamma^2(1-p)} \Big(\frac{s}{t}\Big)^p \int_0^\infty d\alpha_1...d\alpha_4 \delta(\sum \alpha_i-1) \frac{(\alpha_1 \alpha_3)^p(\alpha_2 \alpha_4)^{-p}}{(\sum_{i}\alpha_i(y-x_i)^2 )^4}.
\end{displaymath}
where we have introduced $s=x_{13}^2, t=x_{24}^2$. 
After performing the Wick rotation, the integral over the insertion point $y$ can be readily done,
and we obtain
\begin{align}
I_D(p) = \frac{\pi^{(D-4)/2}\,\Gamma(4-D/2)}{\Gamma^2(1+p)\Gamma^2(1-p)} s^{1+p} t^{1-p}\int d\alpha \delta(\sum \alpha_i-1) \frac{(\alpha_1 \alpha_3)^p(\alpha_2 \alpha_4)^{-p}}{(\alpha_1 \alpha_3 t+\alpha_2 \alpha_4 s)^{4-D/2}}.
\end{align}
Now we introduce new variables (see e.g. \cite{Smirnov:2006ry})
$\alpha_1=\eta_1 \zeta_1,\alpha_2=\eta_1(1-\zeta_1),\alpha_3=\eta_2 \zeta_2$ and $\alpha_4=\eta_2(1-\zeta_2)$. The Jacobian is simply $\eta_1 \eta_2$ and the delta-function constraint implies $\eta_1+\eta_2=1$. The integration over $\eta_1, \eta_2$ can be easily done and we are left with
\begin{displaymath}
I_{D=4-2\epsilon}=\frac{\pi^{-\epsilon}\Gamma^2(-\epsilon)\Gamma(2+\epsilon)}{\Gamma(-2\epsilon)\Gamma^2(1+p)\Gamma^2(1-p)}s^{1+p} t^{1-p}\int_0^1 d\zeta_1 d\zeta_2 \frac{(\zeta_1 \zeta_2)^p((1-\zeta_1)(1-\zeta_2))^{-p}}{(s \zeta_1\zeta_2 +t (1-\zeta_1)(1-\zeta_2))^{2+\epsilon}}\,.
\end{displaymath}
We can separate the $s,t$ dependence by using the Mellin-Barnes representation
\begin{align}
\frac{1}{(X+Y)^\lambda} = \frac{1}{\Gamma(\lambda)} \frac{1}{2\pi i} \int_{\gamma-i\infty}^{\gamma+i\infty} dz \frac{Y^z}{X^{\lambda+z}}\Gamma(\lambda+z)\Gamma(-z)
\end{align}
and performing the integration over $\zeta_1$ and $\zeta_2$. We are left with
\begin{displaymath}
I_{D=4-2\epsilon}=\frac{\sin^2(p\pi)}{(\pi p)^2 \pi^\epsilon s^{2+\epsilon}}\frac{s^{1+p} t^{1-p}}{\Gamma(-2\epsilon)}\frac{1}{2\pi i} \int_{\gamma-i\infty}^{\gamma+i\infty} dz \left(\frac{t}{s}\right)^z \Gamma^2(1-p+z)\Gamma(2+z+\epsilon)\Gamma(-z)\Gamma^2(-1+p-z-\epsilon)
\end{displaymath}
The contour of integration has to be chosen such that all the poles of $\Gamma(...+z)$ are to the left and the poles of $\Gamma(...-z)$ are to the right. We see that the contour is `trapped' between the poles of $\Gamma^2(1-p+z)$ and $\Gamma^2(-1+p-z-\epsilon)$. After analytically continuing the contour (and thereby picking up a residue), we can take the limit $\eps \to 0$.
We obtain
\begin{align}
I_{D=4-2\epsilon}(p)= \frac{\sin \pi p}{\pi p} \frac{4}{\epsilon^2}+{\cal O}(\epsilon^{-1})
\end{align}
The limit $p\rightarrow 0$ exactly reproduces the divergence of the massless scalar box function, as expected.

\section{Three-loop cusp anomalous dimension from integration over $F^{(2)}$.}\label{appendix:asymptotic_exp}
Here we give details of the evaluation of the integral over the insertion point in eq. (\ref{integral_insertion1}).
We found it technically useful to rewrite eq.~(\ref{Ftwoloop}) in terms of 
the more general class of harmonic polylogarithms \cite{Remiddi:1999ew},
\begin{align}\label{F2harmonic}
F^{(2)}(x) =& -\frac{1}{8} \Big[
24 \zeta_2 H_{-1,-1}(x)-12 \zeta_2 H_{-1,0}(x)+24 \zeta_2 H_{0,0}(x)-4
   H_{-2,0,0}(x) \nonumber \\&+8 H_{-1,-1,0,0}(x)-4 H_{-1,0,0,0}(x)+12
   H_{0,0,0,0}(x)-12 \zeta_2 H_{-2}(x)\nonumber \\&+8 \zeta_3 H_{-1}(x)-4 \zeta_3
   H_{0}(x)+107 \zeta_4 \Big]\,.
\end{align}
This has the advantage that it is straightforward to make the logarithmic 
dependence of $F^{(2)}$ manifest. 
We have (e.g. using the algorithm implemented in ref. \cite{Maitre:2005uu}) 
\begin{align}\label{F2_logs_manifest}
F^{(2)}(x) =& -\frac{1}{8} \Big\{ \frac{1}{2} \log^4 x + 12 \zeta_2 \log^2 x-4 \zeta_3  \log x +107 \zeta_4  \\
& \hspace{-1cm} + \log^3 x\left[ -\frac{2}{3} H_{-1}(x) \right] \nonumber \\
&\hspace{-1cm}+ \log^2 x\left[  4 H_{-1,-1}(x) \right] \nonumber \\
&\hspace{-1cm}+ \log x\left[  -8 H_{-2,-1}(x)-8 H_{-1,-2}(x)-12 \zeta_2 H_{-1}(x)+4 H_{-3}(x) \right]  \nonumber \\
&\hspace{-1cm}+ \left[ 24 \zeta_2 H_{-1,-1}(x)+8 H_{-3,-1}(x)+8 H_{-2,-2}(x)+8 H_{-1,-3}(x)+8 \zeta_3 H_{-1}(x)-8
   H_{-4}(x)  \right]  \Big\} \,. \nonumber
\end{align}
In order to perform the integration over the insertion point, we proceed as follows.
First, we can generate any logarithm from powers of $x$ by differentiating formula (\ref{power_int}) w.r.t. $p$.
Second, we use the expansions of the harmonic polylogarithms encountered above in power series around $x=0$.
We have (see \cite{Remiddi:1999ew}) 
\begin{align}\label{equation:HPLs_series}
H_{-n}(x) =& - \sum_{i=1}^{\infty} \frac{(-1)^i x^i}{i^n} \,,  \\
H_{-1,0,-1}(x) =&  \sum_{i=1}^{\infty} \frac{(-1)^i x^i}{i} S_{2}(i) -  \sum_{i=1}^{\infty} \frac{(-1)^i x^i}{i^3}  \,, \\
H_{-1,-1}(x) =&  \sum_{i=1}^{\infty} \frac{(-1)^i x^i}{i} S_{1}(i) -  \sum_{i=1}^{\infty} \frac{(-1)^i x^i}{i^2}  \,, \\
H_{0,-1,-1}(x) =&  \sum_{i=1}^{\infty} \frac{(-1)^i x^i}{i^2} S_{1}(i) -  \sum_{i=1}^{\infty} \frac{(-1)^i x^i}{i^3}  \,.
\end{align}
Here $S_{p}(n)= \sum_{i=1}^{n} 1/i^p$.

Following these steps, we see that
the first line of eq.~(\ref{F2_logs_manifest}) gives a contribution of $56\zeta_{4}$ to $\Gamma^{(3)}_{\rm cusp}$.
Moreover, the second, third, and forth lines contribute $18\zeta_4 , -2 \zeta_4$ and $-6\zeta_4$, respectively,
while the last line does not contribute.
Combining these formulas we straightforwardly obtain the result quoted in the main text.

\section{Evaluating Mellin-Barnes integrals.}\label{appendix:MBintegrals}
In this section, we explain how to evaluate one of Mellin-Barnes integrals that arise in this paper. In section 3, we introduced Mellin-Barnes representation of the two loop answer. Employing the conformal invariance of the answer, we were left with a number of one-fold and one two-fold Mellin-Barnes integrals. The two-fold integral is 
\begin{align}\label{equation:twofold-MBintegral}
& I(x)= \int \frac{dz_1 dz_2}{(2 \pi i)^2} \, x^{-1-z_1} \Gamma^2(-z_1) \Gamma^2(1 + z_1)\Gamma(-z_2) \Gamma(1 + z_2)\,\nonumber \\
& \quad\quad\quad\quad\quad\quad\quad \times \Gamma(1+z_1-z_2) \Gamma(-1-z_1+z_2)(\Psi_{0}( z_2)+\gamma_E)\,,
\end{align}
with $-1<Re(z_1) <Re(z_2)<0$, and $\Psi_{n}$ is the polygamma function such that $\Psi_{n} = \partial^{(n+1)}_{z} \log \Gamma(z)$. This integral can be reduced to one-fold integral with use of the first Barnes lemma with a little twist described in section 3. Taking particular care to separate right and left poles we obtain
\begin{align}\label{equation:onefold-MBintegral}
& I(x)= -\frac{1}{6} \int_{-\frac{1}{2}-i\infty}^{-\frac{1}{2}+i\infty} \frac{dz_1}{2 \pi i}  x^{-1 - z_1} \Gamma(-1 - z_1) \Gamma^2(-z_1) \Gamma^2(1 + z_1) \,\nonumber\\
& \quad\quad\quad\quad \times\Big(6 \Gamma(1 + z_1) \big(2 \gamma_E + \Psi_{0}(-1 - z_1) + \Psi_{0}(1 + z_1)\big) \,\nonumber\\
& \quad\quad\quad\quad\quad + \Gamma(2 + z_1) \big(\pi^2 -6\gamma_E^2 - 12 \gamma_{E} \Psi_{0}(2 + z_1) - 6 \Psi_{0}^2(2 + z_1) - 6 \Psi_{1}(2 + z_1)\big)\Big)
\end{align}                                                                                                                                                                      
\indent In the rest of this appendix, we will present how to evaluate this integral, as an example. $I(x)$ can be expanded in asymptotic series in the limit of $x\ll1$, by the well known procedure of closing the contour. Due to the factor of $x^{-1 - z_1}$ we close the contour on the left hand side of the complex plane. The first pole of the integrand from the right gives the leading contribution in the limit $x\ll1$, the second gives the next-to-leading, etc. By summing residues corresponding to that series of poles we obtain the value of the integral for any value of x.\\
\indent In order to extract those residues, we first simplify the integrand in eq.~(\ref{equation:onefold-MBintegral}) with use of simple gamma function identities
\begin{align}\label{equation:onefold-MBintegral_simplified}
 & I(x)= \int_{-\frac{1}{2}-i\infty}^{-\frac{1}{2}+i\infty} \frac{dz_1}{2 \pi i}  \dfrac{\pi ^3 x^{-1-z_1} {\rm Csc^3}(\pi  z_1)}{6 (1+z_1)} \bigg(6 \big(2\gamma_E (1 + z1)-1\big)  \Psi_0(1+z_1) - 6 \Psi_0(-1-z_1) \nonumber \\
 & \qquad\qquad\qquad\qquad\qquad\qquad (1 + z1) \Big(6 \gamma_E^2 - \pi^2 +6 \Psi_0^2 (2+z_1)+6\Psi_1 (2+z_1)\Big)\bigg).
\end{align}
Integrand in eq.~(\ref{equation:onefold-MBintegral_simplified}) has poles at $z_1 \in \mathbb{Z}$ and involves polygamma functions with the following pole structure for $z\to n=0,-1,-2,...$
\begin{align} \label{equation:polygamma_poles}
  \lim_{z_1\to n} \Psi_0(z_1) = & -\frac{1}{(z_1-n)}+\Big(S_1(-n)-\gamma_E\Big)+\Big(S_2(-n)+\zeta(2)\Big)\,(z_1-n) \nonumber\\
 		& +\Big(S_3(-n)-\zeta(3)\Big)(z_1-n)^2+\Big(S_4(-n)+\zeta(4)\Big)\,(z_1-n)^3 +\cO\big((z_1-n)^4\big) \nonumber\\
  \lim_{z_1\to n} \Psi_1(z_1) = &\: \frac{1}{(z_1-n)^2} +\Big(S_2(-n)+\zeta(2)\Big) +2\Big(S_3(-n)-\zeta(3)\Big)\,(z_1-n)\nonumber \\
		& +3\Big(S_4(-n)+\zeta(4)\Big)\,(z_1-n)^2+\cO\big((z_1-n)^3\big). 
\end{align}
With use of eq.~(\ref{equation:polygamma_poles}) we found that the residue of the integrand at $z_1=-1-i$ for $i\geq1$ is
{\small \begin{align}\label{equation:residue}
{\rm Residue}&\big(z_1=-1-k\big) =  \\
&(-x)^{k} \bigg( \dfrac{2}{k^4} -\dfrac{\log(x)}{k^3} +\dfrac{\pi^2}{3\,k^2} +\dfrac{\pi^2 \log(x) + \log^3(x)}{6k}  -\frac{\log^4(x)}{12} - \frac{\log^3(x) S_{1}(k)}{3}\nonumber \\
& +\log^2(x) \Big(S_{2}(k) - S_{1,1}(k) -\frac{\pi^2}{3}\Big) -2\log(x) \Big(S_{3}(k)- S_{2, 1}(k) - S_{1, 2}(k)+ \frac{\pi^2}{3}  S_{1}(k) \Big) \nonumber \\
&+2 S_{4}(k)-2S_{3, 1}(k)-2S_{2, 2}(k)-2 S_{1, 3}(k)-\pi^2 S_{1,1}(k)+\frac{2 \pi^2 S_{2}(k)}{3} + 2 S_{1}(k) \zeta(3)-\frac{7 \pi^4}{45} \bigg).  \nonumber
\end{align}}
In the above expression we use so-called S-series\cite{Vermaseren:1998uu} i.e. $S_{p}(n)= \sum_{i=1}^{n} 1/i^p$ and $S_{p,r}(n)= \sum_{i=1}^{n} S(r)/i^p$, etc. All nested S-series i.e. $S_{p,r}(n)$ in the above expression originate from a product of two S-series, which we simplify with $S_j(n) S_k(n) = S_{j, k}(n) + S_{k, j}(n) - S_{k + j}(n)$. Now, we would like to sum these residues. This requires series expansions of harmonic polylogarithms\footnote{One might notice that we come across an unnecessary complication going between eq.~(\ref{equation:HPLs_series}) and eq.~(\ref{equation:summing_HPLs}). It is because harmonic polylogarithms are naturally defined in terms of Z-series rather then S-series\cite{Remiddi:1999ew}. On the other hand, we use S-series to follow the conventions of \cite{Remiddi:1999ew} and \cite{Vermaseren:1998uu}.}, already briefly mentioned in eq.~(\ref{equation:HPLs_series}), 
{\small \begin{align}\label{equation:summing_HPLs}
 \sum_{i=1}^{\infty} \frac{(-1)^{i + 1} x^i}{i^a} \qquad \;\;= & \,
   \begin{cases}
     H_{-a}(x) \qquad\qquad\qquad\qquad\qquad\qquad\quad\; & \text{for $a>0$}\\
     \frac{x}{1 + x} \qquad\qquad\qquad\qquad\qquad\qquad\quad\; & \text{for $a=0$}
   \end{cases}& \\
 \sum_{i=1}^{\infty} \frac{(-1)^{i + 1} x^i}{i^a} S_b(i) \;\: = & \,
   \begin{cases}
     -H_{-a, -b}(x) + H_{-a - b}(x) \qquad\qquad\qquad & \text{for $a>0$}\\
     \frac{1}{1 + x} H_{-b}(x) \qquad\qquad\qquad & \text{for $a=0$}
   \end{cases}\nonumber\\
 \sum_{i=1}^{\infty} \frac{(-1)^{i + 1} x^i}{i^a} S_{b, c}(i) = & \,
   \begin{cases}
     H_{-a, -b, -c}(x) - H_{-a, -b - c}(x) - H_{-a - b, -c}(x) + H_{-a-b-c}(x) \quad & \text{for $a>0$}\\
     -\frac{1}{1 + x}H_{-b, -c}(x) + \frac{1}{1 + x} H_{-b - c}(x) \quad & \text{for $a=0$.}
   \end{cases}\nonumber
\end{align}}
Summing the expression in eq.~(\ref{equation:residue}) from $i=1$ to $i=\infty$ using eq.~(\ref{equation:summing_HPLs}) and adding the residue at $z_1=-1$, we found that the integral in eq.~(\ref{equation:twofold-MBintegral}) in terms of harmonic polylogarithms is
\begin{align}
&  I(x)= \dfrac{x \log^4(x)}{12\,(1 + x)} -\dfrac{(x-1) \log^3(x)}{6\,(1 + x)}\, H_{-1}(x) +\dfrac{\log^2(x)}{6\,(1 + x)} \Big(2 \pi^2 x - 6 H_{-1, -1}(x)\Big)\nonumber \\
& -\dfrac{\log(x)}{6\,(1 + x)} \Big( 6 (1 - x) H_{-3}(x) + (x - 3) \pi^2 H_{-1}(x)- 12 H_{-2, -1}(x) - 12 H_{-1, -2}(x)\Big) \nonumber \\
& \qquad + \frac{2 (1-x) H_{-4}(x)}{1 + x} - \frac{2 H_{-3, -1}(x)}{1 + x} - \frac{2 H_{-2, -2}(x)}{1 + x} - \frac{2 H_{-1, -3}(x)}{1 + x} \nonumber \\
& \qquad - \frac{\pi^2 x H_{-2}(x)}{3 (1 + x)} - \frac{\pi^2 H_{-1, -1}(x)}{1 + x} - \frac{2 \zeta(3) H_{-1}(x)}{1 + x} + \frac{\pi^4 (-11 + 17 x)}{180 (1 + x)}.\nonumber 
\end{align}

\end{document}